\documentclass[aps,prb,reprint,superscriptaddress]{revtex4-2}
\usepackage{graphicx}
\usepackage{amsmath}
\usepackage{amssymb}
\usepackage{color}
\begin{document}
\title{First-principles demonstration of Roman surface topological multiferroicity}
\author{Ziwen Wang}
\affiliation{School of Physics, Southeast University, Nanjing 211189, China}
\author{Yisheng Chai}
\email{Email: yschai@cqu.edu.cn}
\affiliation{Low Temperature Physics Laboratory and Chongqing Key Laboratory of Soft Condensed Matter Physics and Smart Materials, College of Physics, Chongqing University, Chongqing 400044, China}
\author{Shuai Dong}
\email{Email: sdong@seu.edu.cn}
\affiliation{School of Physics, Southeast University, Nanjing 211189, China}
\date{\today}

\begin{abstract}
The concept of topology has been widely applied to condensed matter, going beyond the band crossover in reciprocal spaces. A recent breakthrough suggested unconventional topological physics in a quadruple perovskite TbMn$_3$Cr$_4$O$_{12}$, whose magnetism-induced polarization manifests a unique Roman surface topology [Nat. Commun. \textbf{13}, 2373 (2022)]. However, the available experimental evidence based on tiny polarizations of polycrystalline samples is far from sufficient. Here, this topological multiferroicity is demonstrated by using density functional theory calculations, which ideally confirms the Roman surface trajectory of magnetism-induced polarization. In addition, an alternative material in this category is proposed to systematically enhance the performance, by promoting its magnetism-induced polarization to an easily detectable level.
\end{abstract}
\maketitle

\textit{Introduction}.
Topology, as a mathematic concept, has been recognized as an essential ingredient in condensed matter physics since the discovery of quantum Hall effect. Nowadays topological materials with nontrivial electronic/phononic/photonic bands have formed one of the mainstays of quantum materials. And the concept of topology (in both the real and reciprocal spaces) has been gradually and widely infiltrating into other branches of condensed matter, e.g. the magnetic skyrmions \cite{Nagaosa:Nn}. 

In addition, the idea of topology has been applied to the ferroelectric domains in both hexagonal manganites/ferrites and PbTiO$_3$/SrTiO$_3$ superlattices, namely the $\mathbb{Z}_2\times\mathbb{Z}_3$ vortex in the former and the polar vortex/skyrmion/bimeron in the latter \cite{Choi:Nm,Yadav:Nat}. These progresses have greatly pushed forward the understanding of complex physical phenomena in quantum materials, based on elegant principles of mathematics.

Multiferroics, which combine polarity and magnetism in the same phase, have attracted great attentions in the past two decades for their promising magnetoelectric effects \cite{Cheong:Nm,Dong:AP,Lu2019,Vaz2010}. The cross-control between magnetic and polar degrees of freedom is not only highly interesting in physics but also essential for device applications as sensors and storage. The past studies have revealed diversified origins of magnetoelectricity in multiferroics, most of which rely on the spin-orbit coupling (SOC) and/or spin-lattice coupling \cite{Dong2019}. However, the concept of topology has been rarely touched in magnetoelectricity of multiferroics.

\begin{figure}
\includegraphics[width=0.49\textwidth]{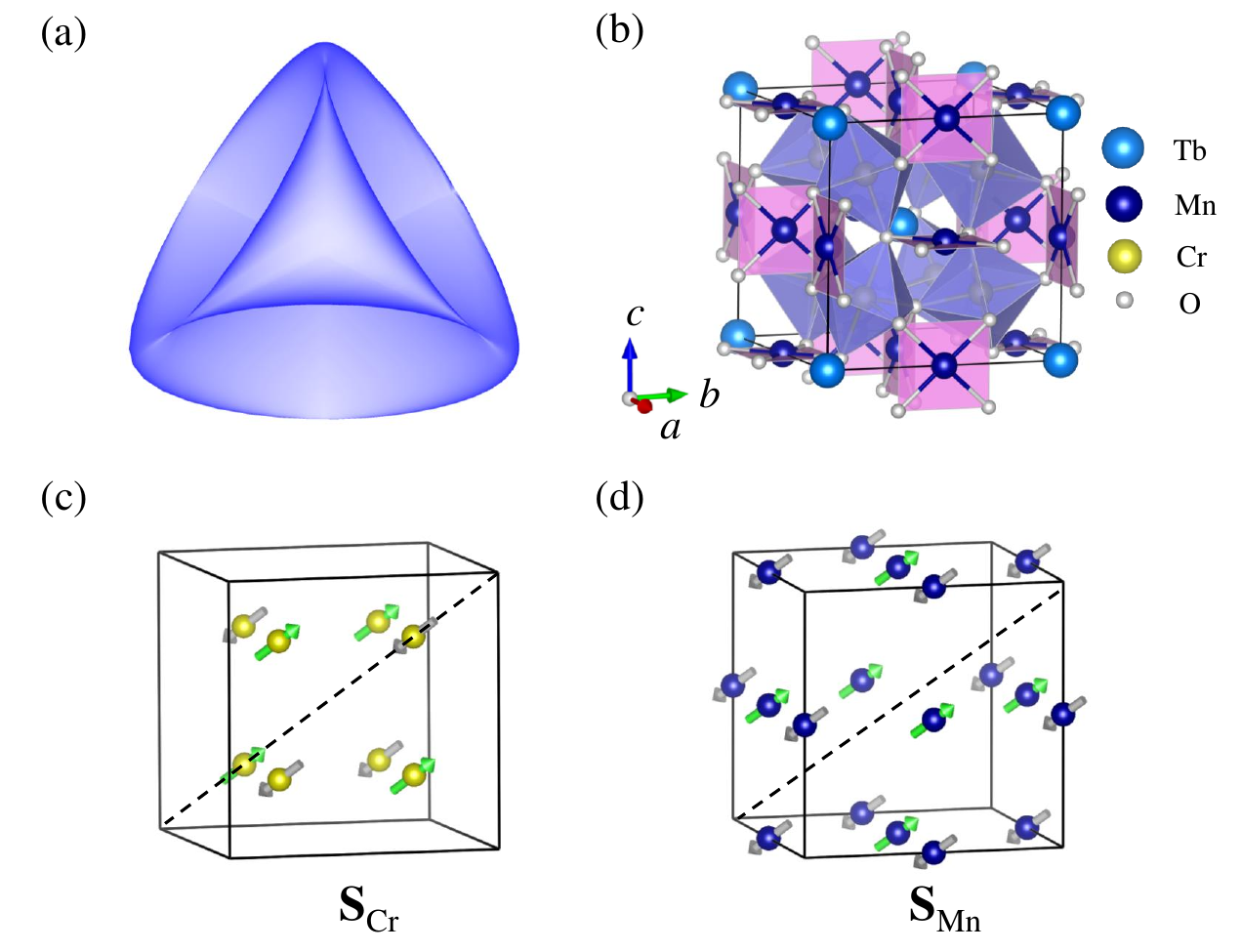}
\caption{(a) Schematic of a Roman surface, which is a quartic non-orientable surface and obtained by sewing a M\"obius strip to the edge of a disk \cite{Liu2022}. (b) Crystal structure of TbMn$_3$Cr$_4$O$_{12}$ with the space group of $Im\bar3$. (c-d) Collinear G-type antiferromagnetic structures of Cr's ($\textbf{S}_{\rm Cr}$'s) and Mn's sublattices ($\textbf{S}_{\rm Mn}$'s), respectively. In the ground state, all spins are pointing along the [111] direction.}
\label{F1}
\end{figure}

Recently, a new brand of topology, the so-called Roman surface, was proposed as a mathematic manifestation of magnetoelectricity in multiferroic $R$Mn$_3$Cr$_4$O$_{12}$ ($R$: rare earth) \cite{Liu2022}. As shown in Fig.~\ref{F1}(a), the Roman surface can be described as:
\begin{equation}
x^2y^2+x^2z^2+y^2z^2-\delta xyz=0,
\label{eq1}
\end{equation}
where $\delta$ is a constant and ($x$, $y$, $z$) denotes a position in the Cartesian coordinate. Roman surface is a kind of non-orientable surface. Its geometric topology can be characterized by a topological invariant orientability number $\omega$: $\omega=0$ if the surface is orientable and $\omega$=1 if the surface is non-orientable. The famous M\"obius strip is a component of the Roman surface, which is well-known for its non-orientable properties. Based on the cubic $Im\bar3$ symmetry of crystal structure [Fig.~\ref{F1}(b)] and double G-type collinear antiferromagnetic order [Fig.~\ref{F1}(c-d)], a high-order magnetism-induced polarization $\textbf{P}$=($P_x$, $P_y$, $P_z$) can plot a trajectory of Roman surface when rotating the spin orientations of Mn$^{3+}$ and Cr$^{3+}$, which renders a topological magnetoelectricity conceputally different from those known ones based on the inverse Dzyaloshinskii-Moriya interaction, exchange striction, or spin-dependent $p$-$d$ hybrizations \cite{Katsura2005,Mostovoy2006,Sergienko:Prb,Sergienko2006-2,Murakawa2010,Dong:AP}.

In Liu \textit{et al.}'s work \cite{Liu2022}, they performed experimental measurements on TbMn$_3$Cr$_4$O$_{12}$ polycrystals. The modulation period of polarization is indeed half of the rotating magnetic field, which is an expected evidence of Roman surface magnetoelectricity. However, such an experimental proof is very preliminary and indirect, only necessary but far from sufficient. The polycrystal samples, without the information of crystalline orientation, is very disadvantage to study the Roman surface. However, most quadruple perovskites can only be synthesized under high-pressure conditions, making the single crystals unavailable at present. Another drawback is its too weak polarization ($\sim100$ $\mu$C/m$^2$ or less), which makes the precise description and potential applications rather challenging.

In this Letter, more quadruple perovskites including TbMn$_3$Cr$_4$O$_{12}$ are investigated by density functional theory (DFT) calculations to circumvent these technical difficulties. The trajectories of induced polarization as a function of spin rotation are obtained, which unambiguously confirm the magnetoelectric topological Roman surface. Our calculations also reveal the induced polarization is a second-order effect of SOC. Based on this fact, we propose a strategy to seek for larger polarizations in this category.

\begin{table}
	\centering
	\caption{The magnetism-induced dipoles $\textbf{D}$'s (in units of $10^{-5}$ e\AA) per u.c. (without lattice relaxation) for spin oritentations along the diagonal directions. The spin configurations are characterized by the $\textbf{S}_{\rm Cr}$-$\textbf{S}_{\rm Mn}$ pairs, as indicated in Fig.~\ref{F1}(c-d). For each antiferromagnetic domain ($\alpha$ or $\beta$), four different dipole orientations are generated by rotating the $\textbf{S}_{\rm Cr}$-$\textbf{S}_{\rm Mn}$ pair synchronously. Another four spin configurations in each domain are not shown for simplify, which own the identical dipoles with their $180^\circ$ counterparts, i.e. $[$111$]$/$[$111$]$ \textit{vs} $[$-1-1-1$]$/$[$-1-1-1$]$. In contrast, the individual $180^\circ$ spin flipping of $\textbf{S}_{\rm Cr}$ (or $\textbf{S}_{\rm Mn}$) will reverse the sign of dipole, forming different antiferromagnetic domains. The equivalent polarization projection along the $x$/$y$/$z$ axis ($P_x$/$P_y$/$P_z$) is $1.7$ $\mu$C/m$^2$.}
	\begin{tabular*}{0.48\textwidth}{@{\extracolsep\fill}lrrrclrrr}
		\hline \hline
		$\textbf{S}_{\rm Cr}$/$\textbf{S}_{\rm Mn}$ & $D_x$ & $D_y$ & $D_z$ &  & $\textbf{S}_{\rm Cr}$/$\textbf{S}_{\rm Mn}$ & $D_x$ & $D_y$ & $D_z$ \\
		\hline
		Domain $\alpha$ & & & & & & & & \\
		\hline
		$[$111$]$/$[$111$]$ & 4 & 4 & 4 &   & $[$11-1$]$/$[$11-1$]$ & -4 & -4 & 4  \\                               
		$[$1-11$]$/$[$1-11$]$ & -4 & 4 & -4 &  & $[$-111$]$/$[$-111$]$ & 4 & -4 & -4 \\
		\hline
		Domain $\beta$ & & & & & & & &\\
		\hline
		$[$111$]$/$[$-1-1-1$]$& -4 & -4 & -4 &   & $[$11-1$]$/$[$-1-11$]$& 4 & 4 & -4  \\                               
		$[$1-11$]$/$[$-11-1$]$ & 4 & -4 & 4 &  & $[$-111$]$/$[$1-1-1$]$ & -4 & 4 & 4 \\
		\hline \hline
	\end{tabular*}
	\label{Tab-1}
\end{table}

\textit{Magnetoelectric Roman surface}.
The magnetism-induced polarization in cubic quadruple perovskites was first observed in LaMn$_3$Cr$_4$O$_{12}$ \cite{Wang2015}, which was rather surprising in such a high symmetric lattice. Its weak polarization ($\sim15$ $\mu$C/m$^2$ in polycrystal samples) made the tiny polar distortion undetectable in the synchrotron X-ray diffraction. Such a polarization was qualitatively reproduced in DFT calculations, although the DFT values were one order of magnitude smaller ($3-7$ $\mu$C/m$^2$ along the [111] direction in single crystal) \cite{Wang2015,Feng2016}.

Therefore, a reliable verification of the topological multiferrocity in $R$Mn$_3$Cr$_4$O$_{12}$ becomes urgently needed, and the gap between the materials experiment and the topological theory can be bridged via precise DFT calculations. According to the previous neutron scattering \cite{Wang2015}, both the Cr$^{3+}$ and Mn$^{3+}$ sublattices are in the G-type antiferromagnetic order, and all spins orient to the [111] direction, as shown in Figs.~\ref{F1}(c-d). Such a magnetic structure forms a polar magnetic point group $31'$ \cite{Liu2022}, which breaks the spatial inversion symmetry.

Our DFT calculation indeed confirms the group theory analysis: the calculated polarization of TbMn$_3$Cr$_4$O$_{12}$ is nonzero with SOC enabled (and zero without SOC). With the rigid high symmetric structure (nonpolar $Im\bar3$), the polarization is $2.8$ $\mu$C/m$^2$, purely from the bias of electronic clouds. After relaxing all ionic positions (with SOC enabled), the net polarization is amplified to $11.9$ $\mu$C/m$^2$. These values are similar to those of sister compound LaMn$_3$Cr$_4$O$_{12}$ \cite{Wang2015}, one order of magnitude below the experimental value ($\sim33$ $\mu$C/m$^2$ for polycrystalline sample) \cite{Liu2022}.

As required by the magnetic point group, the induced dipoles must be parallel (or antiparallel) with the spin orientation, when spins point to the diagonal directions of the cube, i.e. in the ground state (as demonstrated later). This binding relationship between dipoles and spins has been verified in our DFT calculation, as summarized in Table~\ref{Tab-1}. In short, the synchronous $180^\circ$ spin flipping of Cr and Mn sublattices will not change the orientation of dipole. In contrast, the individual $180^\circ$ spin flipping of Cr (or Mn) sublattice will reverse the sign of dipole. These behaviors imply that the magnetism-induced dipole should be in the form of $\textbf{S}_{\rm Cr}\cdot\textbf{S}_{\rm Mn}$-like expression, consistent with the previous proposal \cite{Feng2016}.

\begin{figure*}
\includegraphics[width=0.98\textwidth]{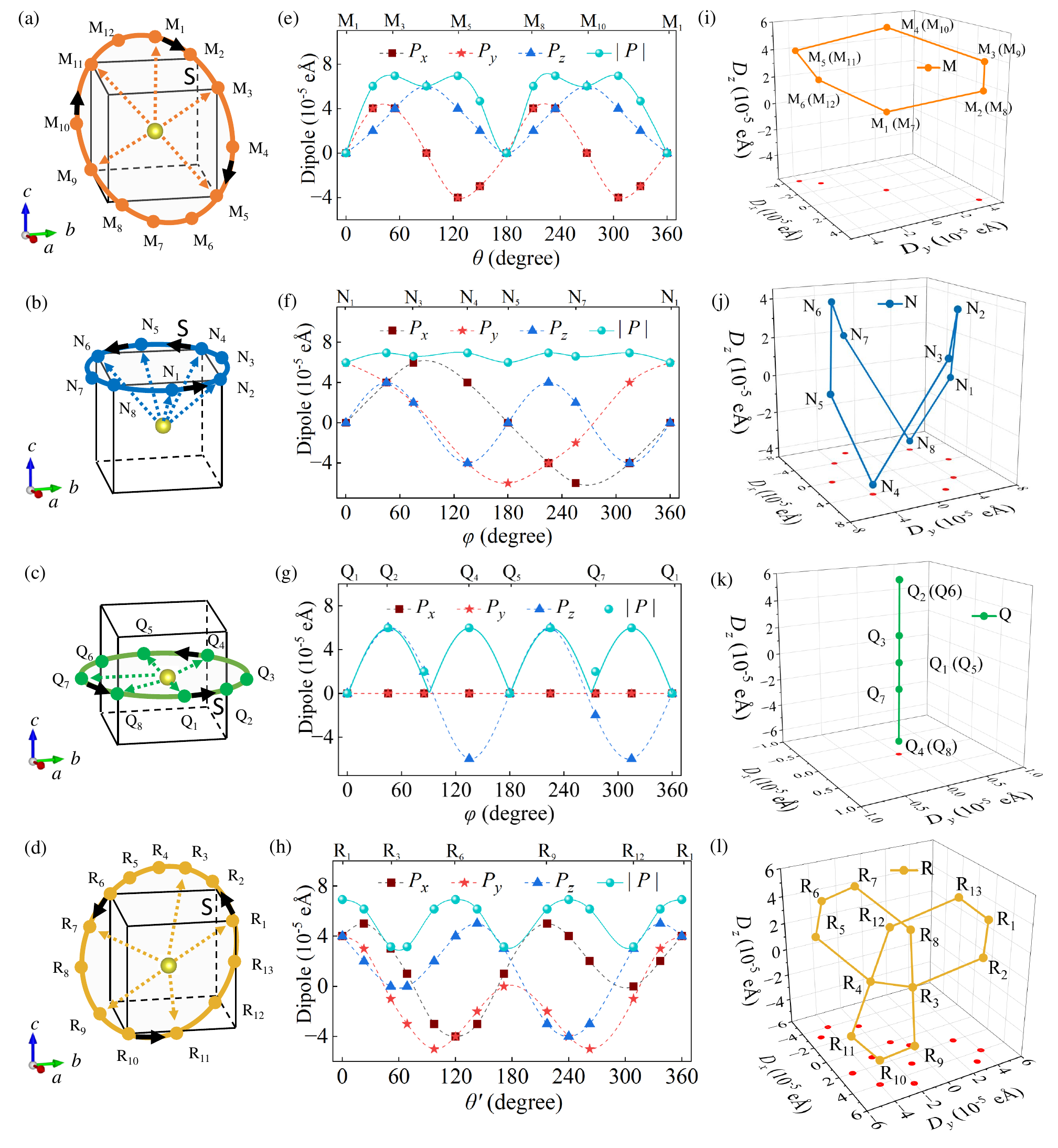}
\caption{Selected closed loops of spin rotation pathes and the induced dipoles. Sampling points in these closed loops are denoted by M, N, Q, and R, respectively. To exclude possible inaccuracy from structural relaxation, here all these spin rotations are done with the strict high symmetric structure. (a-d) Four typical closed loops of spin rotation. (a) Longitudinal rotation (polar angle $\theta\in[0^\circ,180^\circ]$) with fixed azimuthal angles $\varphi=45^\circ$ (and $225^\circ$). (b-c) Latitudinal rotation (azimuthal angle $\phi\in[0^\circ,360^\circ]$) with $\theta=54.7^\circ$ and $\theta=90^\circ$, respectively. (d) Rotation in the ($1\bar{1}1$) plane spanned by three corner points. (e-h) The evolutions of $x$/$y$/$z$ components of dipole in the corresponding rotation loops of spin axis. (i-l) The corresponding dipole trajectories of sampling points in the three-dimensional space and their projections to the $xy$-plane (red dots). (i) and (k) For the trajectories of M and Q sampling points, the two half-periods give rise to the identical trajectory. (i) The elliptic trajectory in three dimensions and the linear projection onto the $xy$-plane. (j) The saddle-like trajectory and the circular projection. (k) The vertical line trajectory and the projection at the original point. (l) The trefoil trajectory in the (111) plane. The projection to the $xy$-plane is also a trefoil one. All these trajectories just fall onto the topological Roman surface.}
\label{F2}
\end{figure*}

\begin{figure*}
	\includegraphics[width=0.95\textwidth]{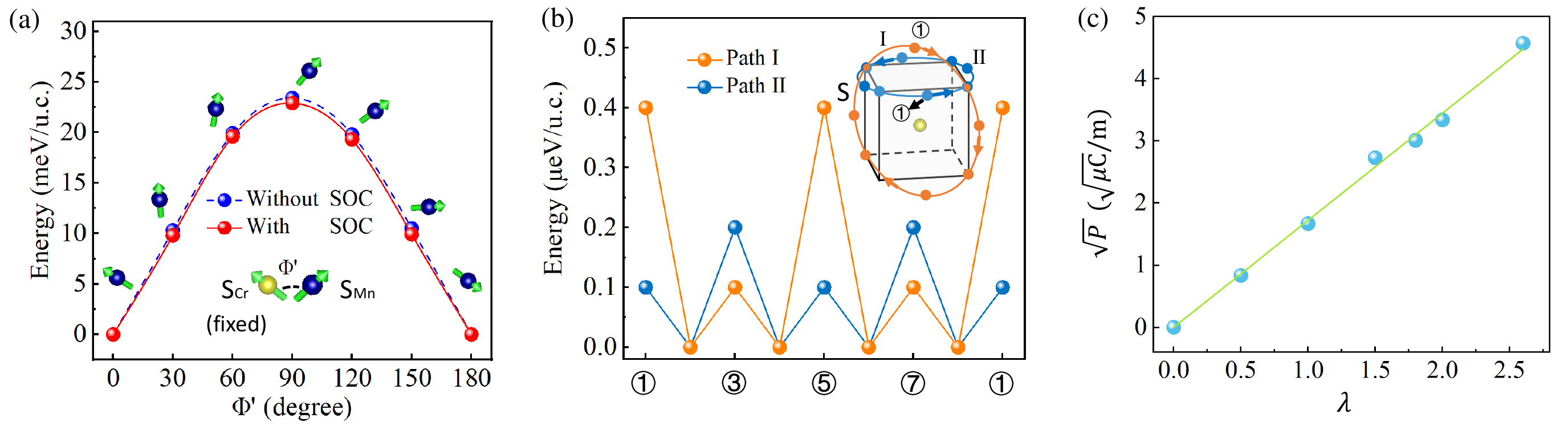}
	\caption{(a) Spin rotation dependence of the individual sublattice. Here $\textbf{S}_{\rm Cr}$ is fixed along the diagonal one and $\textbf{S}_{\rm Mn}$ is rotated, as indicated in the inset. The energy barrier curves with/without SOC almost overlap, implying the dominant contribution is not from SOC. Since the angle ${\Phi'}$ with $0$ ($180$) degree corresponds to the ferroelectric $+P$ ($-P$) state, this energy barrier also can be considered as the theoretical upper limit of ferroelectric switching barrier. (b) Synchronous rotation dependence of two spin sublattices. The rotation path is indicated in the inset. (c) The linear relationship between the square root of induced polarization and the magnitude of SOC.} 
	\label{F3}
\end{figure*}

Then the topological Roman surface of induced polarization can be verified by rotating spins' orientation globally. The whole magnetic structure is represented by a N\'eel vector $\textbf{S}$. Several rotation paths along high symmetric axes are tested.  Taking Fig.~\ref{F2}(a) for example, when spin $\textbf{S}$ rotates longitudinally for a great circle of the sphere, the induced dipole rotates twice [Fig.~\ref{F2}(e)]. The three-dimensional trajectory of dipole vector forms a vertical ellipse, whose projection onto the $xy$-plane is a line [Fig.~\ref{F2}(i)]. Also, the spin rotating direction for a given path (clockwise or anticlockwise) dertermines the moving direction of the dipole, e.g. the clockwise (anticlockwise) spin direction in Fig.~\ref{F2}(a), the anticlockwise (clockwise) dipole direction in Fig.~\ref{F2}(i). More other trajectories can also be found in Fig.~\ref{F2}. As expected, all these trajectories of induced dipoles just fall onto the Roman surface, namely the three components of $\textbf{D}$ strictly fullfill the mathematic expression Eq.~\ref{eq1}, e.g. the topological multiferroicity.

Another vital question is that whether the spins of Cr$^{3+}$ sublattice are always parallel or antiparallel to Mn$^{3+}$ sublattice, especially during the spin rotation? This issue is nontrivial but has not been verified in the experiment~\cite{Liu2022}. If the magnetic interaction between these two superlattices can be described using the Heisenberg spin model, these two sublattices should be completely decoupled, since for each Mn (Cr) the summation over all next-neighboring eight Cr (four Mn) spins is zero. If so, the applied magnetic field can not rotate these two lattices synchronously. Instead, each sublattice can rotate independently and thus the realization of topological Roman surface will be unavailable.

To answer this question, the spin rotation of individual sublattice is calculated with the other sublattice fixed to the [111] orientation. The DFT energy is calculated as a function of spin rotation angle. As shown in Fig.~\ref{F3}(a), the energy reaches the lowest one when the Cr/Mn sublattices are parallel or antiparallel, i.e. the ground state. The engery barrier reaches $24$ meV/u.c. (a non-negligible value) when the spins of Cr/Mn sublattices are perpendicular to each other. And this barrier is almost independent on the SOC, and thus it should be attributed to nonrelativistic interaction between Cr and Mn spin sublattices. Such an energy barrier locks the N\'eel vectors $\textbf{S}_{\rm Cr}$ and $\textbf{S}_{\rm Mn}$ to be parallel or antiparallel in the region of low energy excitation, which is beyond the Heisenberg spin model.

For comparison, the energy fluctuations of synchronous rotation of two spin sublattices are shown in Fig.~\ref{F3}(b). First, the $[$111$]$ orientations are indeed the magnetic easy axes. Second, the energy fluctuations during the spin rotating process are in the order of $0.1-0.4$ $\mu$eV/u.c., which is very small and thus accessible by moderate magnetic fields. Such a weak magnetic anisotropy is reasonable considering for the Mn$^{3+}$'s $3d^4$ and Cr$^{3+}$'s $3d^3$ orbital configurations (to be discussed later). 

In short, the spin pairs can easily rotate in the synchronous manner and the associated polarization draws a Roman surface. The two antiferromagnetic domains lead to two independent Roman surfaces, which are separated by the relative high-energy barrier [Fig.~\ref{F3}(a)] of antiferromagnetic domain walls (i.e., the noncollinear spin texture). As a consequence of such a $\mathbb{Z}_2$-type topology, the $180^\circ$ flip of polarization is forbidden for each Roman surface (i.e., each antiferromagnetic domain) in the low-energy excitation region [Fig.~\ref{F3}(b)], similar to the forbiddance of $180^\circ$ back scattering in the edge state transport of $\mathbb{Z}_2$ topological insulators. 

\textit{Strengthen the polarization}.
Despite the proof of topological magnetoelectricity, there remains an awkward problem, namely its too tiny polarization, which restricts its pratical values. Our following part will try to reveal the underlying physical mechanism and figure out a solution to enhance its performance.

First, to study the relationship between the polarization and the magnitude of SOC, the SOC coefficient ($\lambda$) is artificially tuned \cite{yan2015,Kim2017}. As shown in Fig.~\ref{F3}(c), the square root of polarization grows linearly with increasing SOC, implying that the induced polarization is in proportional to $\lambda^2$. On one hand, such a term is a higher order effect of SOC than the common inverse Dzyaloshinskii-Moriya interaction ($\sim\lambda$) \cite{Sergienko:Prb}, leading to a much weaker induced polarization than those in many other type-II multiferroics \cite{Dong:AP}. On the other hand, the $\lambda^2$ item will allow a drastic enhancement of polarization if larger SOC are involved. 

\begin{figure}
	\includegraphics[width=0.49\textwidth]{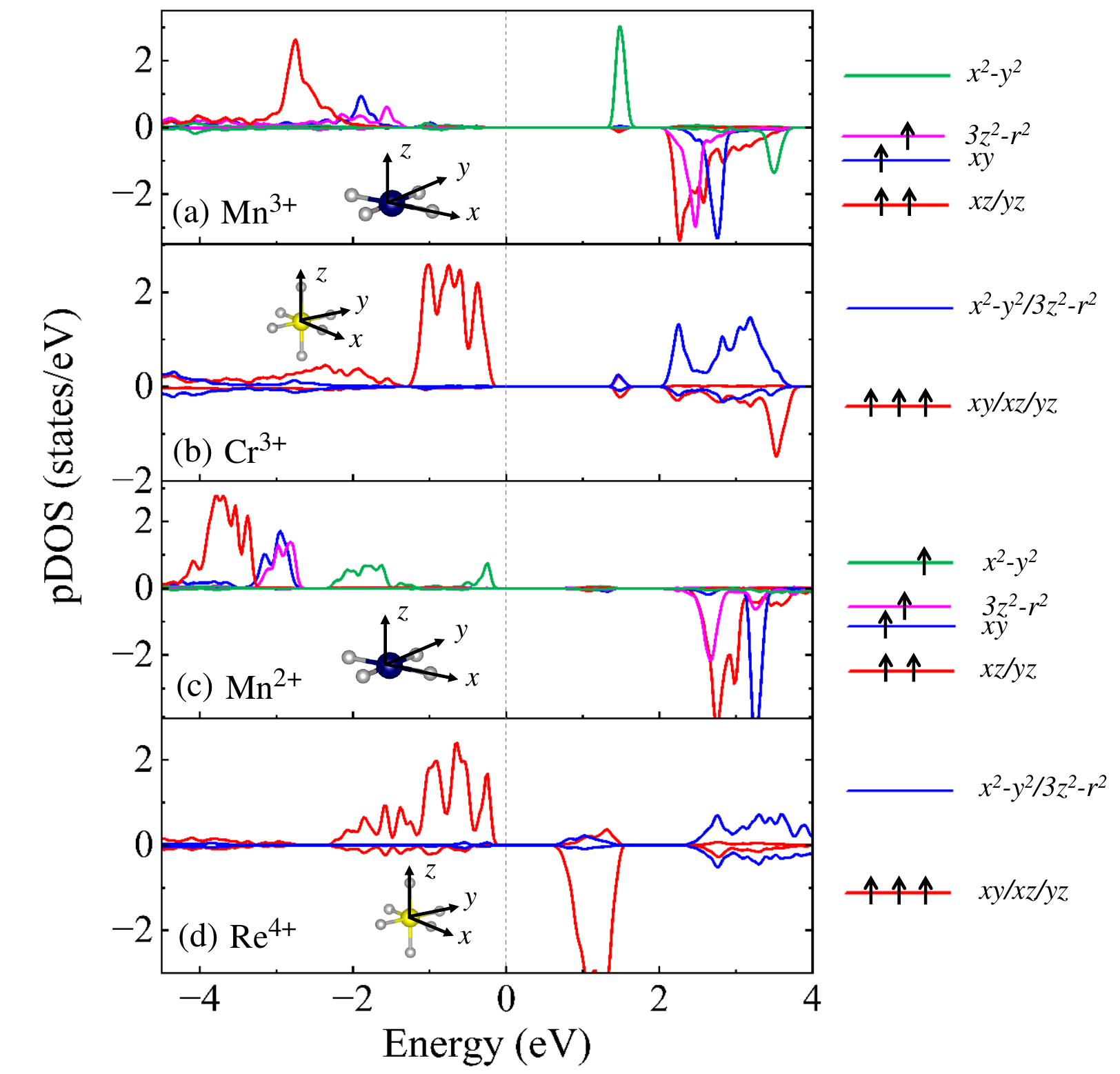}
	\caption{Atomic-orbital projected density of states. (a) Mn$^{3+}$'s $3d$ and (b) Cr$^{3+}$'s $3d$ orbitals of TbMn$_3$Cr$_4$O$_{12}$. (c) Mn$^{2+}$'s $3d$ and (d) Re$^{4+}$'s $5d$ orbitals of BaMn$_3$Re$_4$O$_{12}$. The more itinerant Re's $5d$ electrons give rise to a larger bandwidth and smaller energy gap. Right insets: the corresponding crystal field splitings and the orbital fillings. The electronic configuration in (c-d) can lead to much stronger SOC effects then that in (a-b).}
	\label{F4}
\end{figure}

Next, the electronic structure of TbMn$_3$Cr$_4$O$_{12}$ is analysized, as shown in Fig.~\ref{F4}(a-b). It is clear that both Mn$^{3+}$ and Cr$^{3+}$ are in their high-spin states. For Mn$^{3+}$ ($d^4$), the Fermi level is between the occupied $|3z^2-r^2 \uparrow\rangle$ and empty  $|x^2-y^2 \uparrow\rangle$, which is SOC inactive in the first order level (namely $\langle x^2-y^2 \uparrow|\textbf{L}\cdot\textbf{S}|3z^2-r^2 \uparrow\rangle=0$). Thus, the SOC effect in Mn$^{3+}$ is natural weak. For Cr$^{3+}$ ($d^3$), the Fermi level is between the occupied $t_{\rm 2g}$ and empty $e_{\rm g}$, which allows the SOC effect, as can be seen from our SOC matrix elements analysis in Eqs.~1-4 of SM ~\cite{sm,Weng:Prb}. Since the ferroelectricity arises from the combined action of Cr$^{3+}$ and Mn$^{3+}$ sublattices, the cancellation of the first order SOC in Mn$^{3+}$ originally hinders the large polarization. Thus, to strength the performance of topological magnetoelectricity, the necessary route is to enhance the SOC, especially that on A'-site.

However, although there are many quadruple perovskites with heavier elements (such as Nb, Mo, Tc, Ta, W, Re) occupied in the A'/B sites, our calculations find that in most cases these candidates own different magnetic ground states or become metallic (as summarized in Table~S3 of SM \cite{sm}), similar to earlier experimental studies \cite{li2015,Wu2019,Shiro2013}. Even though, an exceptional candidate BaMn$_3$Re$_4$O$_{12}$ is found to possess the double G-type antiferromagnetic order and insulating properties. After the structural optimization, the space group of BaMn$_3$Re$_4$O$_{12}$ reduces to $I23$ (No. 197). Namely the first-layer Re triangle shrinks towards the diagonal axis and the second-layer Re triangle expands slightly, as shown in Fig.~S2 in SM~\cite{sm}, which break the symmetry of three mirror planes perpendicular to the $a$, $b$, and $c$ axes. Even though, the induced polarization of BaMn$_3$Re$_4$O$_{12}$ still obeys the Roman surface topology (Table~S4 in SM \cite{sm}). 

The most significant result is that the induced polarization of BaMn$_3$Re$_4$O$_{12}$ reaches $810$ $\mu$C/cm$^2$ (and $270$ $\mu$C/cm$^2$ from pure electronic contribution) in our DFT calculation, $68$ times of TbMn$_3$Cr$_4$O$_{12}$. This polarization is comparable to that of the most famous type-II multiferroic TbMnO$_3$ \cite{Kimura2003}, which should be much easier to be precisely detected in experiments and more valuable for practical applications.  Another consequence of strengthening SOC is the increased magnetocrystalline anisotropic energy in BaMn$_3$Re$_4$O$_{12}$, which reaches $\sim2$ meV/u.c. as shown in Fig.~S3 in SM \cite{sm}.

The origin of such a large $P$ can be traced back to its electronic structure, as shown in Fig.~\ref{F4}(c-d). First, the Re's $5d$ orbitals own much larger SOC coefficient than Cr's $3d$ ones, which enhances the SOC interaction in B-site. Second, here the Fermi level lies between the spin-up $t_{\rm 2g}$ and spin-down $e_{\rm g}$ levels for Mn$^{2+}$ ($d^5$), which also systematically enhances the effective SOC interaction in Mn-site as explained above. Third, the band gap becomes smaller in BaMn$_3$Re$_4$O$_{12}$, beneficial to the effective SOC interaction, which needs the virtual hopping between occupied and empty bands. In addition, numerical values of SOC coupling matrices are shown in Fig.~S4 in SM \cite{sm}, which can fully support above analysis.

\textit{Conclusion}.
The Roman surface topological multiferroicity in cubic quadruple perovskites has been demonstrated by first-principles calculation. The low-energy rotation of spin orientation leads to a non-orientable Roman surface for the trajectory of magnetism-induced polarization. As a topological effect, the $180^\circ$ polariztion flip is forbiden for a given Roman surface (antiferromagnetic domain). To overcome the weakness of its too small polarization, another quadruple perovskite BaMn$_3$Re$_4$O$_{12}$ has been predicted to possess the same topological multiferroicity but a much larger polarization. Our work not only confirms and but also extends the Roman surface topological multiferroicity.

\begin{acknowledgments}
This work was supported by National Natural Science Foundation of China (Grant Nos. 11834002 \& 11974065) and the Big Data Computing Center of Southeast University.
\end{acknowledgments}

\bibliography{reference-TM}

\begin{thebibliography}{28}%
\makeatletter
\providecommand \@ifxundefined [1]{%
 \@ifx{#1\undefined}
}%
\providecommand \@ifnum [1]{%
 \ifnum #1\expandafter \@firstoftwo
 \else \expandafter \@secondoftwo
 \fi
}%
\providecommand \@ifx [1]{%
 \ifx #1\expandafter \@firstoftwo
 \else \expandafter \@secondoftwo
 \fi
}%
\providecommand \natexlab [1]{#1}%
\providecommand \enquote  [1]{``#1''}%
\providecommand \bibnamefont  [1]{#1}%
\providecommand \bibfnamefont [1]{#1}%
\providecommand \citenamefont [1]{#1}%
\providecommand \href@noop [0]{\@secondoftwo}%
\providecommand \href [0]{\begingroup \@sanitize@url \@href}%
\providecommand \@href[1]{\@@startlink{#1}\@@href}%
\providecommand \@@href[1]{\endgroup#1\@@endlink}%
\providecommand \@sanitize@url [0]{\catcode `\\12\catcode `\$12\catcode
  `\&12\catcode `\#12\catcode `\^12\catcode `\_12\catcode `\%12\relax}%
\providecommand \@@startlink[1]{}%
\providecommand \@@endlink[0]{}%
\providecommand \url  [0]{\begingroup\@sanitize@url \@url }%
\providecommand \@url [1]{\endgroup\@href {#1}{\urlprefix }}%
\providecommand \urlprefix  [0]{URL }%
\providecommand \Eprint [0]{\href }%
\providecommand \doibase [0]{https://doi.org/}%
\providecommand \selectlanguage [0]{\@gobble}%
\providecommand \bibinfo  [0]{\@secondoftwo}%
\providecommand \bibfield  [0]{\@secondoftwo}%
\providecommand \translation [1]{[#1]}%
\providecommand \BibitemOpen [0]{}%
\providecommand \bibitemStop [0]{}%
\providecommand \bibitemNoStop [0]{.\EOS\space}%
\providecommand \EOS [0]{\spacefactor3000\relax}%
\providecommand \BibitemShut  [1]{\csname bibitem#1\endcsname}%
\let\auto@bib@innerbib\@empty
\bibitem [{\citenamefont {Nagaosa}\ and\ \citenamefont
  {Tokura}(2013)}]{Nagaosa:Nn}%
  \BibitemOpen
  \bibfield  {author} {\bibinfo {author} {\bibfnamefont {N.}~\bibnamefont
  {Nagaosa}}\ and\ \bibinfo {author} {\bibfnamefont {Y.}~\bibnamefont
  {Tokura}},\ }\href@noop {} {\bibfield  {journal} {\bibinfo  {journal} {Nat.
  Nanotechnol.}\ }\textbf {\bibinfo {volume} {8}},\ \bibinfo {pages} {899}
  (\bibinfo {year} {2013})}\BibitemShut {NoStop}%
\bibitem [{\citenamefont {Choi}\ \emph {et~al.}(2010)\citenamefont {Choi},
  \citenamefont {Horibe}, \citenamefont {Yi}, \citenamefont {Choi},
  \citenamefont {Wu},\ and\ \citenamefont {Cheong}}]{Choi:Nm}%
  \BibitemOpen
  \bibfield  {author} {\bibinfo {author} {\bibfnamefont {T.}~\bibnamefont
  {Choi}}, \bibinfo {author} {\bibfnamefont {Y.}~\bibnamefont {Horibe}},
  \bibinfo {author} {\bibfnamefont {H.~T.}\ \bibnamefont {Yi}}, \bibinfo
  {author} {\bibfnamefont {Y.~J.}\ \bibnamefont {Choi}}, \bibinfo {author}
  {\bibfnamefont {W.}~\bibnamefont {Wu}},\ and\ \bibinfo {author}
  {\bibfnamefont {S.-W.}\ \bibnamefont {Cheong}},\ }\href@noop {} {\bibfield
  {journal} {\bibinfo  {journal} {Nat. Mater.}\ }\textbf {\bibinfo {volume}
  {9}},\ \bibinfo {pages} {253} (\bibinfo {year} {2010})}\BibitemShut {NoStop}%
\bibitem [{\citenamefont {Yadav}\ \emph {et~al.}(2016)\citenamefont {Yadav},
  \citenamefont {Nelson}, \citenamefont {Hsu}, \citenamefont {Hong},
  \citenamefont {Clarkson}, \citenamefont {Schlepüetz}, \citenamefont
  {Damodaran}, \citenamefont {Shafer}, \citenamefont {Arenholz}, \citenamefont
  {Dedon}, \citenamefont {Chen}, \citenamefont {Vishwanath}, \citenamefont
  {Minor}, \citenamefont {Chen}, \citenamefont {Scott}, \citenamefont
  {Martin},\ and\ \citenamefont {Ramesh}}]{Yadav:Nat}%
  \BibitemOpen
  \bibfield  {author} {\bibinfo {author} {\bibfnamefont {A.~K.}\ \bibnamefont
  {Yadav}}, \bibinfo {author} {\bibfnamefont {C.~T.}\ \bibnamefont {Nelson}},
  \bibinfo {author} {\bibfnamefont {S.~L.}\ \bibnamefont {Hsu}}, \bibinfo
  {author} {\bibfnamefont {Z.}~\bibnamefont {Hong}}, \bibinfo {author}
  {\bibfnamefont {J.~D.}\ \bibnamefont {Clarkson}}, \bibinfo {author}
  {\bibfnamefont {C.~M.}\ \bibnamefont {Schlepüetz}}, \bibinfo {author}
  {\bibfnamefont {A.~R.}\ \bibnamefont {Damodaran}}, \bibinfo {author}
  {\bibfnamefont {P.}~\bibnamefont {Shafer}}, \bibinfo {author} {\bibfnamefont
  {E.}~\bibnamefont {Arenholz}}, \bibinfo {author} {\bibfnamefont {L.~R.}\
  \bibnamefont {Dedon}}, \bibinfo {author} {\bibfnamefont {D.}~\bibnamefont
  {Chen}}, \bibinfo {author} {\bibfnamefont {A.}~\bibnamefont {Vishwanath}},
  \bibinfo {author} {\bibfnamefont {A.~M.}\ \bibnamefont {Minor}}, \bibinfo
  {author} {\bibfnamefont {L.~Q.}\ \bibnamefont {Chen}}, \bibinfo {author}
  {\bibfnamefont {J.~F.}\ \bibnamefont {Scott}}, \bibinfo {author}
  {\bibfnamefont {L.~W.}\ \bibnamefont {Martin}},\ and\ \bibinfo {author}
  {\bibfnamefont {R.}~\bibnamefont {Ramesh}},\ }\href@noop {} {\bibfield
  {journal} {\bibinfo  {journal} {Nature (London)}\ }\textbf {\bibinfo {volume}
  {530}},\ \bibinfo {pages} {198} (\bibinfo {year} {2016})}\BibitemShut
  {NoStop}%
\bibitem [{\citenamefont {Cheong}\ and\ \citenamefont
  {Mostovoy}(2007)}]{Cheong:Nm}%
  \BibitemOpen
  \bibfield  {author} {\bibinfo {author} {\bibfnamefont {S.-W.}\ \bibnamefont
  {Cheong}}\ and\ \bibinfo {author} {\bibfnamefont {M.}~\bibnamefont
  {Mostovoy}},\ }\href@noop {} {\bibfield  {journal} {\bibinfo  {journal} {Nat.
  Mater.}\ }\textbf {\bibinfo {volume} {6}},\ \bibinfo {pages} {13} (\bibinfo
  {year} {2007})}\BibitemShut {NoStop}%
\bibitem [{\citenamefont {Dong}\ \emph {et~al.}(2015)\citenamefont {Dong},
  \citenamefont {Liu}, \citenamefont {Cheong},\ and\ \citenamefont
  {Ren}}]{Dong:AP}%
  \BibitemOpen
  \bibfield  {author} {\bibinfo {author} {\bibfnamefont {S.}~\bibnamefont
  {Dong}}, \bibinfo {author} {\bibfnamefont {J.-M.}\ \bibnamefont {Liu}},
  \bibinfo {author} {\bibfnamefont {S.~W.}\ \bibnamefont {Cheong}},\ and\
  \bibinfo {author} {\bibfnamefont {Z.~F.}\ \bibnamefont {Ren}},\ }\href@noop
  {} {\bibfield  {journal} {\bibinfo  {journal} {Adv. Phys.}\ }\textbf
  {\bibinfo {volume} {64}},\ \bibinfo {pages} {519} (\bibinfo {year}
  {2015})}\BibitemShut {NoStop}%
\bibitem [{\citenamefont {Lu}\ \emph {et~al.}(2019)\citenamefont {Lu},
  \citenamefont {Wu}, \citenamefont {Lin},\ and\ \citenamefont {Liu}}]{Lu2019}%
  \BibitemOpen
  \bibfield  {author} {\bibinfo {author} {\bibfnamefont {C.}~\bibnamefont
  {Lu}}, \bibinfo {author} {\bibfnamefont {M.}~\bibnamefont {Wu}}, \bibinfo
  {author} {\bibfnamefont {L.}~\bibnamefont {Lin}},\ and\ \bibinfo {author}
  {\bibfnamefont {J.-M.}\ \bibnamefont {Liu}},\ }\href
  {https://doi.org/10.1093/nsr/nwz091} {\bibfield  {journal} {\bibinfo
  {journal} {Natl. Sci. Rev.}\ }\textbf {\bibinfo {volume} {6}},\ \bibinfo
  {pages} {653} (\bibinfo {year} {2019})}\BibitemShut {NoStop}%
\bibitem [{\citenamefont {Vaz}\ \emph {et~al.}(2010)\citenamefont {Vaz},
  \citenamefont {Hoffman}, \citenamefont {Ahn},\ and\ \citenamefont
  {Ramesh}}]{Vaz2010}%
  \BibitemOpen
  \bibfield  {author} {\bibinfo {author} {\bibfnamefont {C.~A.}\ \bibnamefont
  {Vaz}}, \bibinfo {author} {\bibfnamefont {J.}~\bibnamefont {Hoffman}},
  \bibinfo {author} {\bibfnamefont {C.~H.}\ \bibnamefont {Ahn}},\ and\ \bibinfo
  {author} {\bibfnamefont {R.}~\bibnamefont {Ramesh}},\ }\href@noop {}
  {\bibfield  {journal} {\bibinfo  {journal} {Adv. Mater.}\ }\textbf {\bibinfo
  {volume} {22}},\ \bibinfo {pages} {2900} (\bibinfo {year}
  {2010})}\BibitemShut {NoStop}%
\bibitem [{\citenamefont {Dong}\ \emph {et~al.}(2019)\citenamefont {Dong},
  \citenamefont {Xiang},\ and\ \citenamefont {Dagotto}}]{Dong2019}%
  \BibitemOpen
  \bibfield  {author} {\bibinfo {author} {\bibfnamefont {S.}~\bibnamefont
  {Dong}}, \bibinfo {author} {\bibfnamefont {H.~J.}\ \bibnamefont {Xiang}},\
  and\ \bibinfo {author} {\bibfnamefont {E.}~\bibnamefont {Dagotto}},\
  }\href@noop {} {\bibfield  {journal} {\bibinfo  {journal} {Natl. Sci. Rev.}\
  }\textbf {\bibinfo {volume} {6}},\ \bibinfo {pages} {629} (\bibinfo {year}
  {2019})}\BibitemShut {NoStop}%
\bibitem [{\citenamefont {Liu}\ \emph {et~al.}(2022)\citenamefont {Liu},
  \citenamefont {Pi}, \citenamefont {Zhou}, \citenamefont {Liu}, \citenamefont
  {Shen}, \citenamefont {Ye}, \citenamefont {Qin}, \citenamefont {Mi},
  \citenamefont {Chen}, \citenamefont {Zhao}, \citenamefont {Zhou},
  \citenamefont {Guo}, \citenamefont {Yu}, \citenamefont {Chai}, \citenamefont
  {Weng},\ and\ \citenamefont {Long}}]{Liu2022}%
  \BibitemOpen
  \bibfield  {author} {\bibinfo {author} {\bibfnamefont {G.}~\bibnamefont
  {Liu}}, \bibinfo {author} {\bibfnamefont {M.}~\bibnamefont {Pi}}, \bibinfo
  {author} {\bibfnamefont {L.}~\bibnamefont {Zhou}}, \bibinfo {author}
  {\bibfnamefont {Z.}~\bibnamefont {Liu}}, \bibinfo {author} {\bibfnamefont
  {X.}~\bibnamefont {Shen}}, \bibinfo {author} {\bibfnamefont {X.}~\bibnamefont
  {Ye}}, \bibinfo {author} {\bibfnamefont {S.}~\bibnamefont {Qin}}, \bibinfo
  {author} {\bibfnamefont {X.}~\bibnamefont {Mi}}, \bibinfo {author}
  {\bibfnamefont {X.}~\bibnamefont {Chen}}, \bibinfo {author} {\bibfnamefont
  {L.}~\bibnamefont {Zhao}}, \bibinfo {author} {\bibfnamefont {B.}~\bibnamefont
  {Zhou}}, \bibinfo {author} {\bibfnamefont {J.}~\bibnamefont {Guo}}, \bibinfo
  {author} {\bibfnamefont {X.}~\bibnamefont {Yu}}, \bibinfo {author}
  {\bibfnamefont {Y.}~\bibnamefont {Chai}}, \bibinfo {author} {\bibfnamefont
  {H.}~\bibnamefont {Weng}},\ and\ \bibinfo {author} {\bibfnamefont
  {Y.}~\bibnamefont {Long}},\ }\href
  {https://doi.org/10.1038/s41467-022-29764-w} {\bibfield  {journal} {\bibinfo
  {journal} {Nat. Commun.}\ }\textbf {\bibinfo {volume} {13}},\ \bibinfo
  {pages} {2373} (\bibinfo {year} {2022})}\BibitemShut {NoStop}%
\bibitem [{\citenamefont {Katsura}\ \emph {et~al.}(2005)\citenamefont
  {Katsura}, \citenamefont {Nagaosa},\ and\ \citenamefont
  {Balatsky}}]{Katsura2005}%
  \BibitemOpen
  \bibfield  {author} {\bibinfo {author} {\bibfnamefont {H.}~\bibnamefont
  {Katsura}}, \bibinfo {author} {\bibfnamefont {N.}~\bibnamefont {Nagaosa}},\
  and\ \bibinfo {author} {\bibfnamefont {A.~V.}\ \bibnamefont {Balatsky}},\
  }\href {https://doi.org/10.1103/PhysRevLett.95.057205} {\bibfield  {journal}
  {\bibinfo  {journal} {Phys. Rev. Lett.}\ }\textbf {\bibinfo {volume} {95}},\
  \bibinfo {pages} {057205} (\bibinfo {year} {2005})}\BibitemShut {NoStop}%
\bibitem [{\citenamefont {Mostovoy}(2006)}]{Mostovoy2006}%
  \BibitemOpen
  \bibfield  {author} {\bibinfo {author} {\bibfnamefont {M.}~\bibnamefont
  {Mostovoy}},\ }\href {https://doi.org/10.1103/PhysRevLett.96.067601}
  {\bibfield  {journal} {\bibinfo  {journal} {Phys. Rev. Lett.}\ }\textbf
  {\bibinfo {volume} {96}},\ \bibinfo {pages} {067601} (\bibinfo {year}
  {2006})}\BibitemShut {NoStop}%
\bibitem [{\citenamefont {Sergienko}\ and\ \citenamefont
  {Dagotto}(2006)}]{Sergienko:Prb}%
  \BibitemOpen
  \bibfield  {author} {\bibinfo {author} {\bibfnamefont {I.~A.}\ \bibnamefont
  {Sergienko}}\ and\ \bibinfo {author} {\bibfnamefont {E.}~\bibnamefont
  {Dagotto}},\ }\href@noop {} {\bibfield  {journal} {\bibinfo  {journal} {Phys.
  Rev. B}\ }\textbf {\bibinfo {volume} {73}},\ \bibinfo {pages} {094434}
  (\bibinfo {year} {2006})}\BibitemShut {NoStop}%
\bibitem [{\citenamefont {Sergienko}\ \emph {et~al.}(2006)\citenamefont
  {Sergienko}, \citenamefont {\ifmmode~\mbox{\c{S}}\else \c{S}\fi{}en},\ and\
  \citenamefont {Dagotto}}]{Sergienko2006-2}%
  \BibitemOpen
  \bibfield  {author} {\bibinfo {author} {\bibfnamefont {I.~A.}\ \bibnamefont
  {Sergienko}}, \bibinfo {author} {\bibfnamefont {C.}~\bibnamefont
  {\ifmmode~\mbox{\c{S}}\else \c{S}\fi{}en}},\ and\ \bibinfo {author}
  {\bibfnamefont {E.}~\bibnamefont {Dagotto}},\ }\href
  {https://doi.org/10.1103/PhysRevLett.97.227204} {\bibfield  {journal}
  {\bibinfo  {journal} {Phys. Rev. Lett.}\ }\textbf {\bibinfo {volume} {97}},\
  \bibinfo {pages} {227204} (\bibinfo {year} {2006})}\BibitemShut {NoStop}%
\bibitem [{\citenamefont {Murakawa}\ \emph {et~al.}(2010)\citenamefont
  {Murakawa}, \citenamefont {Onose}, \citenamefont {Miyahara}, \citenamefont
  {Furukawa},\ and\ \citenamefont {Tokura}}]{Murakawa2010}%
  \BibitemOpen
  \bibfield  {author} {\bibinfo {author} {\bibfnamefont {H.}~\bibnamefont
  {Murakawa}}, \bibinfo {author} {\bibfnamefont {Y.}~\bibnamefont {Onose}},
  \bibinfo {author} {\bibfnamefont {S.}~\bibnamefont {Miyahara}}, \bibinfo
  {author} {\bibfnamefont {N.}~\bibnamefont {Furukawa}},\ and\ \bibinfo
  {author} {\bibfnamefont {Y.}~\bibnamefont {Tokura}},\ }\href
  {https://doi.org/10.1103/PhysRevLett.105.137202} {\bibfield  {journal}
  {\bibinfo  {journal} {Phys. Rev. Lett.}\ }\textbf {\bibinfo {volume} {105}},\
  \bibinfo {pages} {137202} (\bibinfo {year} {2010})}\BibitemShut {NoStop}%
\bibitem [{\citenamefont {Wang}\ \emph {et~al.}(2015)\citenamefont {Wang},
  \citenamefont {Chai}, \citenamefont {Zhou}, \citenamefont {Cao},
  \citenamefont {Cruz}, \citenamefont {Yang}, \citenamefont {Dai},
  \citenamefont {Yin}, \citenamefont {Yuan}, \citenamefont {Zhang},
  \citenamefont {Yu}, \citenamefont {Azuma}, \citenamefont {Shimakawa},
  \citenamefont {Zhang}, \citenamefont {Dong}, \citenamefont {Sun},
  \citenamefont {Jin},\ and\ \citenamefont {Long}}]{Wang2015}%
  \BibitemOpen
  \bibfield  {author} {\bibinfo {author} {\bibfnamefont {X.}~\bibnamefont
  {Wang}}, \bibinfo {author} {\bibfnamefont {Y.}~\bibnamefont {Chai}}, \bibinfo
  {author} {\bibfnamefont {L.}~\bibnamefont {Zhou}}, \bibinfo {author}
  {\bibfnamefont {H.}~\bibnamefont {Cao}}, \bibinfo {author} {\bibfnamefont
  {C.-d.}\ \bibnamefont {Cruz}}, \bibinfo {author} {\bibfnamefont
  {J.}~\bibnamefont {Yang}}, \bibinfo {author} {\bibfnamefont {J.}~\bibnamefont
  {Dai}}, \bibinfo {author} {\bibfnamefont {Y.}~\bibnamefont {Yin}}, \bibinfo
  {author} {\bibfnamefont {Z.}~\bibnamefont {Yuan}}, \bibinfo {author}
  {\bibfnamefont {S.}~\bibnamefont {Zhang}}, \bibinfo {author} {\bibfnamefont
  {R.}~\bibnamefont {Yu}}, \bibinfo {author} {\bibfnamefont {M.}~\bibnamefont
  {Azuma}}, \bibinfo {author} {\bibfnamefont {Y.}~\bibnamefont {Shimakawa}},
  \bibinfo {author} {\bibfnamefont {H.}~\bibnamefont {Zhang}}, \bibinfo
  {author} {\bibfnamefont {S.}~\bibnamefont {Dong}}, \bibinfo {author}
  {\bibfnamefont {Y.}~\bibnamefont {Sun}}, \bibinfo {author} {\bibfnamefont
  {C.}~\bibnamefont {Jin}},\ and\ \bibinfo {author} {\bibfnamefont
  {Y.}~\bibnamefont {Long}},\ }\href
  {https://doi.org/10.1103/PhysRevLett.115.087601} {\bibfield  {journal}
  {\bibinfo  {journal} {Phys. Rev. Lett.}\ }\textbf {\bibinfo {volume} {115}},\
  \bibinfo {pages} {087601} (\bibinfo {year} {2015})}\BibitemShut {NoStop}%
\bibitem [{\citenamefont {Feng}\ and\ \citenamefont {Xiang}(2016)}]{Feng2016}%
  \BibitemOpen
  \bibfield  {author} {\bibinfo {author} {\bibfnamefont {J.~S.}\ \bibnamefont
  {Feng}}\ and\ \bibinfo {author} {\bibfnamefont {H.~J.}\ \bibnamefont
  {Xiang}},\ }\href {https://doi.org/10.1103/PhysRevB.93.174416} {\bibfield
  {journal} {\bibinfo  {journal} {Phys. Rev. B}\ }\textbf {\bibinfo {volume}
  {93}},\ \bibinfo {pages} {174416} (\bibinfo {year} {2016})}\BibitemShut
  {NoStop}%
\bibitem [{\citenamefont {Yan}\ \emph {et~al.}(2015)\citenamefont {Yan},
  \citenamefont {Stadtm{\"u}ller}, \citenamefont {Haag}, \citenamefont
  {Jakobs}, \citenamefont {Seidel}, \citenamefont {Jungkenn}, \citenamefont
  {Mathias}, \citenamefont {Cinchetti}, \citenamefont {Aeschlimann},\ and\
  \citenamefont {Felser}}]{yan2015}%
  \BibitemOpen
  \bibfield  {author} {\bibinfo {author} {\bibfnamefont {B.}~\bibnamefont
  {Yan}}, \bibinfo {author} {\bibfnamefont {B.}~\bibnamefont
  {Stadtm{\"u}ller}}, \bibinfo {author} {\bibfnamefont {N.}~\bibnamefont
  {Haag}}, \bibinfo {author} {\bibfnamefont {S.}~\bibnamefont {Jakobs}},
  \bibinfo {author} {\bibfnamefont {J.}~\bibnamefont {Seidel}}, \bibinfo
  {author} {\bibfnamefont {D.}~\bibnamefont {Jungkenn}}, \bibinfo {author}
  {\bibfnamefont {S.}~\bibnamefont {Mathias}}, \bibinfo {author} {\bibfnamefont
  {M.}~\bibnamefont {Cinchetti}}, \bibinfo {author} {\bibfnamefont
  {M.}~\bibnamefont {Aeschlimann}},\ and\ \bibinfo {author} {\bibfnamefont
  {C.}~\bibnamefont {Felser}},\ }\href@noop {} {\bibfield  {journal} {\bibinfo
  {journal} {Nat. Commun.}\ }\textbf {\bibinfo {volume} {6}},\ \bibinfo {pages}
  {10167} (\bibinfo {year} {2015})}\BibitemShut {NoStop}%
\bibitem [{\citenamefont {Kim}\ \emph {et~al.}(2017)\citenamefont {Kim},
  \citenamefont {Kim}, \citenamefont {Wang}, \citenamefont {Sinova},\ and\
  \citenamefont {Wu}}]{Kim2017}%
  \BibitemOpen
  \bibfield  {author} {\bibinfo {author} {\bibfnamefont {J.}~\bibnamefont
  {Kim}}, \bibinfo {author} {\bibfnamefont {K.-W.}\ \bibnamefont {Kim}},
  \bibinfo {author} {\bibfnamefont {H.}~\bibnamefont {Wang}}, \bibinfo {author}
  {\bibfnamefont {J.}~\bibnamefont {Sinova}},\ and\ \bibinfo {author}
  {\bibfnamefont {R.}~\bibnamefont {Wu}},\ }\href
  {https://doi.org/10.1103/PhysRevLett.119.027201} {\bibfield  {journal}
  {\bibinfo  {journal} {Phys. Rev. Lett.}\ }\textbf {\bibinfo {volume} {119}},\
  \bibinfo {pages} {027201} (\bibinfo {year} {2017})}\BibitemShut {NoStop}%
\bibitem [{sm()}]{sm}%
  \BibitemOpen
  \href@noop {} {}\bibinfo {note} {See Supplementary Materials [url] for more
  detials about the methods, related data and discussion, which includes Refs.
  \cite{Kresse1996,Perdew:Prl08,Dudarev1998,King-Smith1993,Weng:Prb}}\BibitemShut
  {NoStop}%
\bibitem [{\citenamefont {Weng}\ \emph {et~al.}(2020)\citenamefont {Weng},
  \citenamefont {Li},\ and\ \citenamefont {Dong}}]{Weng:Prb}%
  \BibitemOpen
  \bibfield  {author} {\bibinfo {author} {\bibfnamefont {Y.}~\bibnamefont
  {Weng}}, \bibinfo {author} {\bibfnamefont {X.}~\bibnamefont {Li}},\ and\
  \bibinfo {author} {\bibfnamefont {S.}~\bibnamefont {Dong}},\ }\href@noop {}
  {\bibfield  {journal} {\bibinfo  {journal} {Phys. Rev. B}\ }\textbf {\bibinfo
  {volume} {102}},\ \bibinfo {pages} {180401(R)} (\bibinfo {year}
  {2020})}\BibitemShut {NoStop}%
\bibitem [{\citenamefont {Li}\ \emph {et~al.}(2015)\citenamefont {Li},
  \citenamefont {Retuerto}, \citenamefont {Deng}, \citenamefont {Sarkar} \emph
  {et~al.}}]{li2015}%
  \BibitemOpen
  \bibfield  {author} {\bibinfo {author} {\bibfnamefont {M.-R.}\ \bibnamefont
  {Li}}, \bibinfo {author} {\bibfnamefont {M.}~\bibnamefont {Retuerto}},
  \bibinfo {author} {\bibfnamefont {Z.}~\bibnamefont {Deng}}, \bibinfo {author}
  {\bibfnamefont {T.}~\bibnamefont {Sarkar}}, \emph {et~al.},\ }\href@noop {}
  {\bibfield  {journal} {\bibinfo  {journal} {Chem. Mater.}\ }\textbf {\bibinfo
  {volume} {27}},\ \bibinfo {pages} {211} (\bibinfo {year} {2015})}\BibitemShut
  {NoStop}%
\bibitem [{\citenamefont {Wu}\ \emph {et~al.}(2019)\citenamefont {Wu},
  \citenamefont {Frank}, \citenamefont {Han}, \citenamefont {Croft},
  \citenamefont {Walker}, \citenamefont {Greenblatt},\ and\ \citenamefont
  {Li}}]{Wu2019}%
  \BibitemOpen
  \bibfield  {author} {\bibinfo {author} {\bibfnamefont {M.}~\bibnamefont
  {Wu}}, \bibinfo {author} {\bibfnamefont {C.~E.}\ \bibnamefont {Frank}},
  \bibinfo {author} {\bibfnamefont {Y.}~\bibnamefont {Han}}, \bibinfo {author}
  {\bibfnamefont {M.}~\bibnamefont {Croft}}, \bibinfo {author} {\bibfnamefont
  {D.}~\bibnamefont {Walker}}, \bibinfo {author} {\bibfnamefont
  {M.}~\bibnamefont {Greenblatt}},\ and\ \bibinfo {author} {\bibfnamefont
  {M.-R.}\ \bibnamefont {Li}},\ }\href@noop {} {\bibfield  {journal} {\bibinfo
  {journal} {Inorg. Chem.}\ }\textbf {\bibinfo {volume} {58}},\ \bibinfo
  {pages} {10280} (\bibinfo {year} {2019})}\BibitemShut {NoStop}%
\bibitem [{\citenamefont {Shiro}\ \emph {et~al.}(2013)\citenamefont {Shiro},
  \citenamefont {Yamada}, \citenamefont {Ikeda}, \citenamefont {Ohgushi},
  \citenamefont {Mizumaki}, \citenamefont {Takahashi}, \citenamefont
  {Nishiyama}, \citenamefont {Inoue},\ and\ \citenamefont
  {Irifune}}]{Shiro2013}%
  \BibitemOpen
  \bibfield  {author} {\bibinfo {author} {\bibfnamefont {K.}~\bibnamefont
  {Shiro}}, \bibinfo {author} {\bibfnamefont {I.}~\bibnamefont {Yamada}},
  \bibinfo {author} {\bibfnamefont {N.}~\bibnamefont {Ikeda}}, \bibinfo
  {author} {\bibfnamefont {K.}~\bibnamefont {Ohgushi}}, \bibinfo {author}
  {\bibfnamefont {M.}~\bibnamefont {Mizumaki}}, \bibinfo {author}
  {\bibfnamefont {R.}~\bibnamefont {Takahashi}}, \bibinfo {author}
  {\bibfnamefont {N.}~\bibnamefont {Nishiyama}}, \bibinfo {author}
  {\bibfnamefont {T.}~\bibnamefont {Inoue}},\ and\ \bibinfo {author}
  {\bibfnamefont {T.}~\bibnamefont {Irifune}},\ }\href@noop {} {\bibfield
  {journal} {\bibinfo  {journal} {Inorg. Chem.}\ }\textbf {\bibinfo {volume}
  {52}},\ \bibinfo {pages} {1604} (\bibinfo {year} {2013})}\BibitemShut
  {NoStop}%
\bibitem [{\citenamefont {Kimura}\ \emph {et~al.}(2003)\citenamefont {Kimura},
  \citenamefont {Goto}, \citenamefont {Shintani}, \citenamefont {Ishizaka},
  \citenamefont {Arima},\ and\ \citenamefont {Tokura}}]{Kimura2003}%
  \BibitemOpen
  \bibfield  {author} {\bibinfo {author} {\bibfnamefont {T.}~\bibnamefont
  {Kimura}}, \bibinfo {author} {\bibfnamefont {T.}~\bibnamefont {Goto}},
  \bibinfo {author} {\bibfnamefont {H.}~\bibnamefont {Shintani}}, \bibinfo
  {author} {\bibfnamefont {K.}~\bibnamefont {Ishizaka}}, \bibinfo {author}
  {\bibfnamefont {T.}~\bibnamefont {Arima}},\ and\ \bibinfo {author}
  {\bibfnamefont {Y.}~\bibnamefont {Tokura}},\ }\href
  {https://doi.org/10.1038/nature02018} {\bibfield  {journal} {\bibinfo
  {journal} {Nature}\ }\textbf {\bibinfo {volume} {426}},\ \bibinfo {pages}
  {55} (\bibinfo {year} {2003})}\BibitemShut {NoStop}%
\bibitem [{\citenamefont {Kresse}\ and\ \citenamefont
  {Furthm\"uller}(1996)}]{Kresse1996}%
  \BibitemOpen
  \bibfield  {author} {\bibinfo {author} {\bibfnamefont {G.}~\bibnamefont
  {Kresse}}\ and\ \bibinfo {author} {\bibfnamefont {J.}~\bibnamefont
  {Furthm\"uller}},\ }\href {https://doi.org/10.1103/PhysRevB.54.11169}
  {\bibfield  {journal} {\bibinfo  {journal} {Phys. Rev. B}\ }\textbf {\bibinfo
  {volume} {54}},\ \bibinfo {pages} {11169} (\bibinfo {year}
  {1996})}\BibitemShut {NoStop}%
\bibitem [{\citenamefont {Perdew}\ \emph {et~al.}(2008)\citenamefont {Perdew},
  \citenamefont {Ruzsinszky}, \citenamefont {Csonka}, \citenamefont {Vydrov},
  \citenamefont {Scuseria}, \citenamefont {Constantin}, \citenamefont {Zhou},\
  and\ \citenamefont {Burke}}]{Perdew:Prl08}%
  \BibitemOpen
  \bibfield  {author} {\bibinfo {author} {\bibfnamefont {J.~P.}\ \bibnamefont
  {Perdew}}, \bibinfo {author} {\bibfnamefont {A.}~\bibnamefont {Ruzsinszky}},
  \bibinfo {author} {\bibfnamefont {G.~I.}\ \bibnamefont {Csonka}}, \bibinfo
  {author} {\bibfnamefont {O.~A.}\ \bibnamefont {Vydrov}}, \bibinfo {author}
  {\bibfnamefont {G.~E.}\ \bibnamefont {Scuseria}}, \bibinfo {author}
  {\bibfnamefont {L.~A.}\ \bibnamefont {Constantin}}, \bibinfo {author}
  {\bibfnamefont {X.}~\bibnamefont {Zhou}},\ and\ \bibinfo {author}
  {\bibfnamefont {K.}~\bibnamefont {Burke}},\ }\href@noop {} {\bibfield
  {journal} {\bibinfo  {journal} {Phys. Rev. Lett.}\ }\textbf {\bibinfo
  {volume} {100}},\ \bibinfo {pages} {136406} (\bibinfo {year}
  {2008})}\BibitemShut {NoStop}%
\bibitem [{\citenamefont {Dudarev}\ \emph {et~al.}(1998)\citenamefont
  {Dudarev}, \citenamefont {Botton}, \citenamefont {Savrasov}, \citenamefont
  {Humphreys},\ and\ \citenamefont {Sutton}}]{Dudarev1998}%
  \BibitemOpen
  \bibfield  {author} {\bibinfo {author} {\bibfnamefont {S.~L.}\ \bibnamefont
  {Dudarev}}, \bibinfo {author} {\bibfnamefont {G.~A.}\ \bibnamefont {Botton}},
  \bibinfo {author} {\bibfnamefont {S.~Y.}\ \bibnamefont {Savrasov}}, \bibinfo
  {author} {\bibfnamefont {C.~J.}\ \bibnamefont {Humphreys}},\ and\ \bibinfo
  {author} {\bibfnamefont {A.~P.}\ \bibnamefont {Sutton}},\ }\href
  {https://doi.org/10.1103/PhysRevB.57.1505} {\bibfield  {journal} {\bibinfo
  {journal} {Phys. Rev. B}\ }\textbf {\bibinfo {volume} {57}},\ \bibinfo
  {pages} {1505} (\bibinfo {year} {1998})}\BibitemShut {NoStop}%
\bibitem [{\citenamefont {King-Smith}\ and\ \citenamefont
  {Vanderbilt}(1993)}]{King-Smith1993}%
  \BibitemOpen
  \bibfield  {author} {\bibinfo {author} {\bibfnamefont {R.~D.}\ \bibnamefont
  {King-Smith}}\ and\ \bibinfo {author} {\bibfnamefont {D.}~\bibnamefont
  {Vanderbilt}},\ }\href {https://doi.org/10.1103/PhysRevB.47.1651} {\bibfield
  {journal} {\bibinfo  {journal} {Phys. Rev. B}\ }\textbf {\bibinfo {volume}
  {47}},\ \bibinfo {pages} {1651} (\bibinfo {year} {1993})}\BibitemShut
  {NoStop}%
\end{thebibliography}%
\bibliographystyle{apsrev4-2}
\end{document}